# Role of the ferromagnetic component in the ferroelectricity of YMnO$_3$

Kiran Singh, Natalia Bellido, Charles Simon, Julien Varignon, Marie-B Lepetit

Laboratoire CRISMAT,

CNRS-ENSICAEN, 14050 Caen, France

Albin De Muer,

Laboratoire National des Champs Magnétiques Intenses

CNRS-Université de Grenoble Joseph Fourier

38042 Grenoble France.

Stephane Pailhès,

Laboratoire Léon Brillouin

CNRS-CEA, CEN Saclay, 91191 Gif/Yvette, France.

LPMCN – University Claude Bernard Lyon I

69622 Villeurbanne Cedex, France

Abstract: We performed magnetic and ferroelectric measurements, first principle calculations and Landau theory analysis on hexagonal YMnO$_3$. The polarization and the AFM order parameter were found to present different temperature dependence at $T_N$. A linear coupling between these two order parameters is thus forbidden in the Landau theory and P6$_3$cm cannot be the magnetic group. The only compatible magnetic group is P6'$_3$. In this group however, Landau theory predicts the possibility of a ferromagnetic component and of a linear coupling between the dielectric constant and the AFM order parameter. On one hand we performed dielectric constant measurements under magnetic field that clearly exhibit a metamagnetic transition, and thus confirm these predictions. On the other hand careful magnetization measurements show a small by non null FM component along the c-axis direction. Finally the Landau analysis within the P6'$_3$ magnetic group shows that only the polarization square is coupled to the magnetic orders and thus neither the magnetization nor the AFM order can be reversed by an applied electric field.

**Introduction**

Hexagonal $YMnO_3$ presents ferroelectricity and antiferromagnetism [1,2] and can be considered as the prototype of type I ferroelectric antiferromagnetic materials in which the details of the magnetoelectric coupling can be studied. In the present paper, we performed magnetic and ferroelectric measurements in order to clarify the situation of this system which is very often taken as an example in the literature. Starting from these new data, we have written the free energy F in a Landau approach and explained the temperature dependence of the different order parameters.

Though $YMnO_3$ was studied for years since the pioneer work of Yakel et al. in 1963, the exact crystalline and magnetic structures are still under debate. The temperature of the ferroelectric transition is still for example not completely clear. Located by some authors at 920K [3], recent X rays measurements propose 1258K [4]. These discrepancies are not fully understood and are possibly due to some changes in the oxygen deficiency when the sample is heated. Despite these discrepancies, we can try to summarize the knowledge of this ferroelectric transition as follows: A transition corresponding to a unit-cell tripling and a change in space group from centrosymmetric $P6_3/mmc$ (#194) to polar $P6_3cm$ (#185) is observed in this temperature range. The symmetric group $P6_3mmc$ reduces to $P6_3cm$ by a rotation of $MnO_5$ polyedra and a displacement of yttrium with respect to manganese atoms along the c axis of the structure [5,6], inducing a c axis polarization. Furthermore, an intermediate phase with the space group $P6_3mcm$ can be derived from group theory [7], however it was not observed in the recent measurements[8]. The authors rather observe some evidence for an isosymmetric phase transition at about 920 K, which involves a sharp decrease in estimated polarization. This second transition correlates with several previous reports of anomalies in physical properties in this temperature region [9], but is not really understood.

Concerning the antiferromagnetic transition, the critical temperature $T_N$ is reported at 74K. The magnetism is arising from $Mn^{3+}$ ions, in configuration $3d^4$, with a spin equal to 2 (high spin). Neutron diffraction measurements [10,11,12] showed that the structure is antiferromagnetic with in ab-plane moments. Following Bertaut et al. [13], Munoz et al. [10], $\Gamma_1$ of the $P6_3cm$ group. More recently, a spin polarized analysis shows that the group is in fact rather $P6_3$ (or $P6_3$') [14].

Concerning the structural determination, a giant magneto elastic coupling was observed by powder neutron diffraction at the magnetic transition suggesting very large atomic displacements induced by the magnetic ordering up to 0.1Å [15] (isostructural transition within the group $P6_3cm$ in this work). Concerning the low temperature polarization and dielectric constant measurements, no real precise macroscopic measurements were reported on a single crystal (such measurements exist in thin films). Not directly related to the structure, but with some incidence on the symmetry, second harmonic generation measurements [16] and inelastic neutron measurements [14] were also reported on this compound.

**Experimental details**

The single crystal we used was grown and characterized in Groningen by G. Nenert from the group of T. Palstra. All the measurements were performed on the same crystal. The sample size for dielectric measurements is a=1.1mm, b=1.5mm and c=0.3mm. Magnetic measurements were performed with a QD MPMS-5 SQUID magnetometer. Dielectric and polarization measurements were performed in a QD PPMS-14 with Agilent 4284A LCR meter and Keithley 6517A respectively. Magnetic field above 14T up to 25T was achieved in LNCMI Grenoble with the same experimental setup as in Caen to measure dielectric constant and their own setup for magnetization. One antiferromagnetic neutron diffraction peak was measured on 4F triple axis spectrometer in Laboratoire Léon Brillouin in Saclay on the same single crystal.

**An antiferromagnetic transition with a Neel transition at 74K**

We performed neutron scattering experiments on a neutron triple axis spectrometer and checked the crystal orientation and crystalline quality. Assuming that the magnetic group can be $P6_3cm$, $P6_3$ or $P6'_3$, the magnetic peak 100 describes the antiferromagnetic order parameter. On fig. 1, the temperature dependence of the amplitude of the 100 antiferromagnetic peak is reported, showing the magnetic transition at $T_N$. On the same figure, we also reported the ab component of the dielectric constant ε, which presents an anomaly at the same temperature (the c component of ε does not present any anomaly at this temperature). The similarity below $T_N$ between the temperature dependence of the antiferromagnetic order parameter and the non linear part of ε suggests that they are closely related.

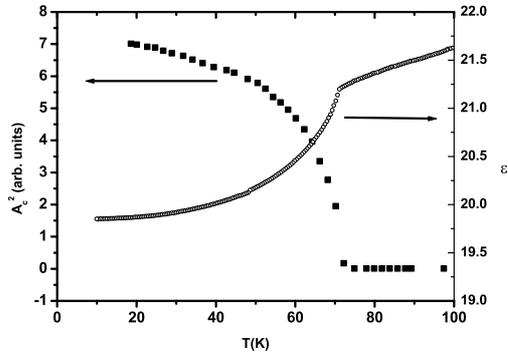

*Figure 1 : Temperature dependence of the intensity of the 101 antiferromagnetic peak (left scale) and the ab component of the dielectric constant (right scale).*

**Polarization measurements**

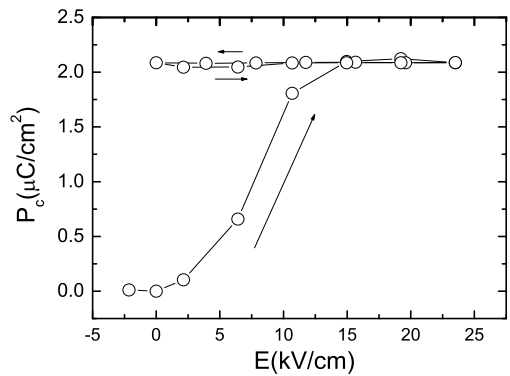

*Figure 2 : The polarization cycle at 30K after cooling in a zero electric field (polarization is very small after this procedure, see figure 4). The second branch of the measurement ensures that the measured current is not due to leakage (polarization is already switched).*

Concerning the temperature dependence of polarization, there is no report in the literature on single crystals. Polarization measurements along c-axis (the only one allowed by symmetry) are shown in fig. 2. The polarization presents a clear hysteresis cycle which is unlikely reported in crystals at low temperature. A polarization value of $2\mu C/cm^2$ is found here at 30K. This value is perfectly compatible with the estimated polarization (**P**$=\Sigma$ **q**$_i$**r**$_i$) obtained from atomic displacements observed in powder neutron diffraction [11]. From the atomic positions determined by neutron diffraction at 10K and 300K [15], we computed the polarization using density functional theory and a Berry phases approach. The calculation was performed using the B1PW hybrid functionals specifically designed for the treatment of ferroelectric oxides [17]. At low temperature (10K), we found a $1\mu C/cm^2$ polarization while at room temperature (300K) the polarization was computed to be $5\mu C/cm^2$ in agreement with the $4.5\mu C/cm^2$ measured value [18].

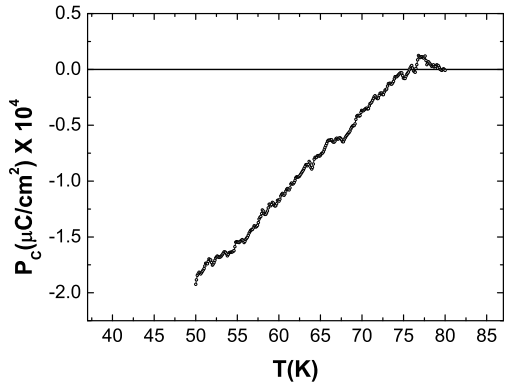

*Figure 3: Temperature dependence of polarization after cooling in zero electric field. The polarization is very small (Note the factor 10 000 which multiplies the polarization here, compared to fully polarized sample see figure 2).*

There are no direct polarization measurements versus temperature reported until now. We tried to fully polarize the single crystal by the application of a large electric field and then to vary the temperature. Unfortunately, this procedure is very difficult in YMnO$_3$ and we never succeeded. Very often, some accident appears before the end of the measurement due to the conductivity of the helium gas or to any other conducting process. For this reason, we designed an alternative procedure. Without full polarization, we ramped, many times, up and down the temperature to measure polarization, and found a small but measurable signal which gives a small but real temperature dependence of the polarization (fig. 3). On this figure, the temperature dependence below $T_N$ is clearly linear, that is entirely different from the antiferromagnetic magnetization (see fig. 1). From their atomic displacements, Lee et al. [15] reported an estimated polarization versus temperature (calculated as $\mathbf{P}=\Sigma q_i \mathbf{r}_i$ from the atomic positions) which is possibly linear in T in their figure, despite the fact that they suggest it can be similar to the order parameter antiferromagnetic order parameter (which is not experimentally found linear in temperature). This is an important point, since a different behavior of the polarization and the antiferromagnetic order parameter at $T_N$ implies the absence of a linear coupling between the two in a Landau theory.

**First Landau analysis**

Let us try to describe this transition in a Landau analysis. The experimental results forbid a linear coupling between the polarization and the antiferromagnetic order parameters. Three magnetic groups are proposed in the literature, namely $P6_3cm$, $P6_3$ and $P6'_3$. The antiferromagnetic order found both by Bertaud [13] and Munoz [10] corresponds to the $\Gamma_1$ irreductible representation of the $P6_3cm$ group. However this order can also be associated with symmetry vectors of the $P6_3$ and $P6'_3$ groups, namely with the first vector $\mathbf{V_1}$ of the $\Gamma_1$ irreductible representation of the $P6_3$ group and with the first vector $\mathbf{V_1}$ of the $\Gamma_4$ irreductible representation of the $P6'_3$ group (see fig. 4).

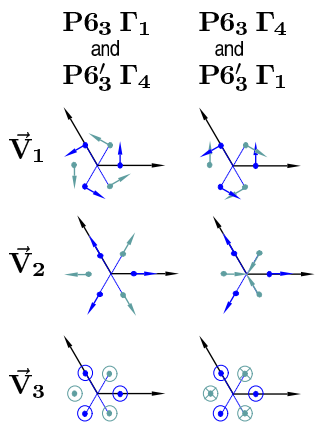

*Figure 4: The order parameters $V_i$ of the magnetic groups $P6_3$ and $P6'_3$ for one irreductible representation. Let us point out that both the $\Gamma_1$ representation of $P6_3$ and the $\Gamma_4$ representation of $P6'_3$ are represented three times with symmetry vectors $V_1$, $V_2$ and $V_3$*

Let us point out that both the $\Gamma_1$ representation of $P6_3$ and the $\Gamma_4$ representation of $P6'_3$ are represented three times with symmetry vectors $\mathbf{V_1}$, $\mathbf{V_2}$ and $\mathbf{V_3}$ (see fig. 4). Remembering that the polarization belong to the $\Gamma_1$ representation in any of the three groups, one sees immediately that both $P6_3cm$ and $P6_3$ groups allow linear coupling between the polarisation and the antiferromagnetic order parameters. The experimental findings are thus only compatible with the $P6'_3$ group since only this group forbids this linear coupling. In this group, the three antiferromagnetic order parameters belong to the $\Gamma_4$ representation, and thus should appear in pairs in the Landau free energy. Let us define these three order parameters.

$$\mathbf{A} = 1/6r \sum_i \mathbf{r_i}_\wedge \mathbf{S_i} = 1/6r \sum_i \mathbf{r_i}_\wedge \mathbf{S_i^{ab}}$$

$$B = 1/6r \sum_i \mathbf{r_i} \cdot \mathbf{S_i} = 1/6r \sum_i \mathbf{r_i} \cdot \mathbf{S_i^{ab}}$$

$$\mathbf{M} = 1/6 \sum_i \mathbf{S_i} = 1/6 \sum_i \mathbf{S_i^c}$$

where the summations over i run over the six Mn atoms of the unit cell. $\mathbf{r_i}$ refer to the in plane components of the Mn atoms position vectors (note that $\sum_i \mathbf{r_i} = \mathbf{0}$ and $\forall i \; |\mathbf{r_i}|=r$), and $\mathbf{S_i}$ to the Mn atomic spins ($\mathbf{S_i} = \mathbf{S_i^{ab}} + \mathbf{S_i^c}$ where $\mathbf{S_i^{ab}}$ is the in-plane component of the Mn spins and $\mathbf{S_i^c}$ is the c axis component). $\mathbf{A}$ and $\mathbf{M}$ are vectors along the **c** direction while B is a scalar. $\mathbf{A}$ is associated with the $\mathbf{V_1}$ symmetry vector and is called the toroidal order parameter [19], B is associated with $\mathbf{V_2}$ and corresponds to the in plane divergence of the spins and $\mathbf{M}$ is associated with $\mathbf{V_3}$ and correspond to a magnetization along the **c** axis (see fig. 4). If one notes $A = S^{ab} \cos\phi$ and $B = S^{ab} \sin\phi$, the intensity of the 100 magnetic peak is proportional to the square of $S^{ab}$ whatever is the angle $\phi$. At this stage of the discussion, it is not possible to chose among the different possibilities and we will keep $S^{ab}$ as the only order parameter of the antiferromagnetism.

Let us first only consider the antiferromagnetic and the polarization orders. In the paramagnetic state, i.e. for $T > T_N$, $S^{ab} = 0$, but **P** is not zero. This is one of the important issues of this compound. The Landau free energy can thus be expressed as

$$F = a_2(T-T_N)(S^{ab})^2 + a_4(S^{ab})^4 - \alpha_2(P^2-P_0^2) + \alpha_4(P^4-P_0^4)$$
$$+ \gamma (S^{ab})^2(P^2-P_0^2) \qquad (1)$$

where $a_2$, $a_4$, $\alpha_2$, $\alpha_4$, $\gamma$ are the temperature independent Landau expansion coefficients. The first two parameters (terms in $S^{ab}$) correspond to the antiferromagnetic energy. The $P^2$ and $P^4$ terms are the usual terms for a ferroelectric material. In order to get F=0 at the antiferromagnetic phase transition $T_N$, we subtracted a constant contribution to F to account for the non null polarization at $T_N$ : $P_0=P(T_N)$. The term $\gamma (S^{ab})^2(P^2-P_0^2)$ corresponds to the coupling between the antiferromagnetic order and the polarization. It should be noticed that this term exists whatever the symmetry of the system since it involves only squares of the order parameters.

Expression (1) of the free energy is derived with respect to $S^{ab}$ and P to find the minimum ($t=T_N-T$):

$\partial F/\partial S^{ab} = 2S^{ab} [-a_2 t + 2a_4(S^{ab})^2 + \gamma(P^2-P_0^2)] = 0$

$\partial F/\partial P = 2P [-\alpha_2 + 2\alpha_4 P^2 + \gamma(S^{ab})^2] = 0$.

At $T_N$ this leads to

$P_0^2 = \alpha_2/2\alpha_4$

and below $T_N$ this leads to

$S^{ab} = [2a_2\alpha_4 t/(4a_4\alpha_4 - \gamma^2)]^{1/2}$

$P^2 = P_0^2 [1 - \gamma/\alpha_2\ 2\alpha_4 a_2 t/(4\alpha_4 a_4 \gamma^2)]$

If one compares these results to the experimental data, one can notice that the polarization varies linearly in t at the magnetic transition as experimentally observed (see fig. 3). This analysis also predicts that the critical shape of $(S^{ab})^2$ versus t should be linear, that of a classical second order phase transition. In fact, as it is for most magnetic phase transitions, the higher order terms in the free energy make the temperature dependence over a large scale of temperature different from the mean field prediction. Here for example, the best fit is a power law in $t^{1/3}$ (not shown in fig. 1).

Coming back to the anomaly of the dielectric constant at $T_N$, one gets using the second derivative of F with respect to P

$1/\chi_e = d^2F/dP^2 = 2[-\alpha_2 + 6\alpha_4 P^2 + \gamma(S^{ab})^2] = 4[\alpha_2 + \gamma(S^{ab})^2]$

where $\chi_e$ is the dielectric susceptibility.

$$\varepsilon = 1+\chi_e = 1+1/[4\alpha_2 + 4\gamma (S^{ab})^2] \approx 1+1/4\alpha_2 - (\gamma/4\alpha_2) (S^{ab})^2$$

for small values of $\gamma/\alpha_2$.

If one compares this result to the experimental data of fig.1, the Landau analysis predicts that the critical shape of $\varepsilon$ versus t should be similar that of $(S^{ab})^2$, as can be seen figure 1.

**Measurements under magnetic field**

This analysis predicts a coupling between the dielectric constant and the antiferromagnetic order parameter. A classical method to observe the antiferromagnetism consists in searching for the metamagnetic transition. Magnetization M versus magnetic field H presents a clear linear behavior up to 23T without any anomaly whatever the temperature from 5K to 300K (not shown). Despite the complete absence of anomaly in M(H) up to 23T (due to the weakness of the ferromagnetic component), we performed the measurement of $\varepsilon(H)$ under the same conditions (magnetic field along the **c** axis) (fig. 5a) and found a temperature dependent anomaly. We reported these anomalies and obtained a phase diagram (fig. 5b) which is characteristic of an antiferromagnetic compound under magnetic field. This technique allows us to obtain this antiferromagnetic phase diagram for the first time. It also proves for the first time the existence of a coupling between polarization and magnetism in $YMnO_3$.

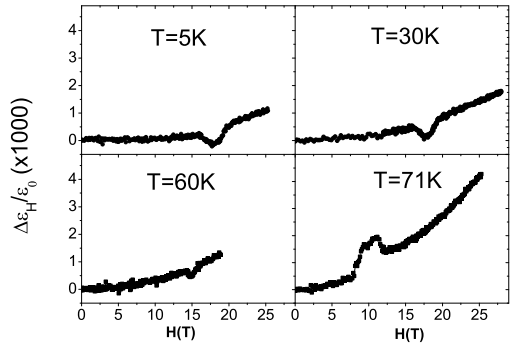

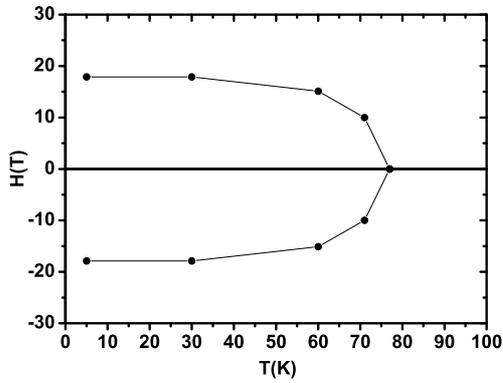

*Figure 5 : Magnetic field dependence of the ab dielectric constant (H is along the c axis) and the corresponding phase diagram (H,T).*

In addition, we measured polarization versus magnetic field. Since this effect is also expected to be very small, we used the same procedure as for temperature dependence (ramping many times the magnetic field from -14T to +14T and extracting the periodic signal from the raw data). One can see on figure 6 the anomaly at the metamagnetic transition reported on figure 5 by dielectric measurements. The existence of a coupling between the magnetic component and polarization is then fully confirmed.

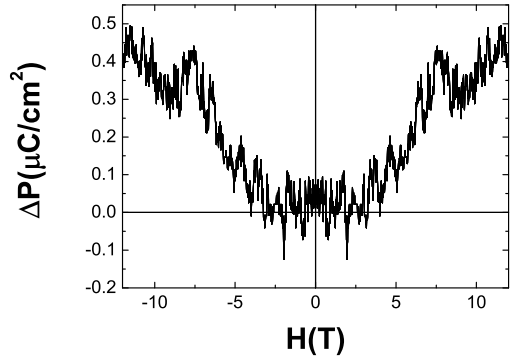

*Figure 6: Magnetic field dependence of the polarization at 71K.*

**A ferromagnetic compound with a critical transition at $T_N$**

In the previous Landau analysis, the possible ferromagnetic component (symmetry vector $V_3$) was neglected, however if the magnetic group is correct a ferromagnetic c-axis component should exist. Let us note that such a component was suggested by Bertaud in 1960 [13] and more recently by Pailhes et al. [14]. In order to evidence this ferromagnetic contribution we performed precise magnetic measurements on the SQUID magnetometer at low magnetic field. The sample was cooled down from 100K (above $T_N$) to 10K either under an applied magnetic field along the c axis of the crystal (Field Cooled =FC) or without any field (Zero Field Cooled = ZFC). After cooling, the magnetization was always measured under an applied field. This procedure, assuming that the applied field is too small to reverse the magnetization, evidences the ferromagnetic component. On fig. 7, one can see the result of this procedure. It clearly exhibits a ferromagnetic c axis component. On this figure, the applied magnetic field is 0.1T, which corresponds to the saturation of the induced moment (the magnetic dependence is not shown). In addition, the temperature dependence of this ferromagnetic component --- as obtained by the subtraction of FC and ZFC curves (inset of fig. 7) --- is very different from the antiferromagnetic temperature dependence, showing that these two order parameters are not linearly coupled. The absolute value of the magnetic moment ($10^{-5} \mu_B$ per atom) has to be compared to the 4 $\mu_B$ per atom of the antiferromagnetic order.

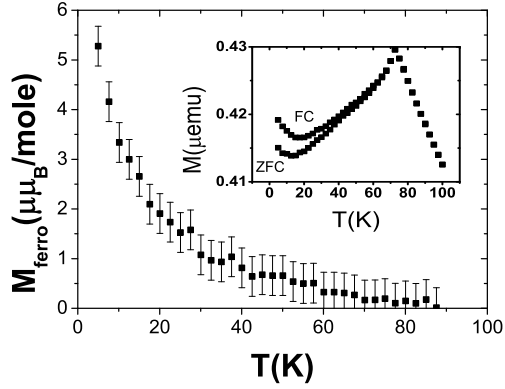

*Figure 7: The ferromagnetic component M as function of temperature. In the inset, the raw data in field cooled (FC) and zero field cooled (ZFC) modes.*

**Second Landau analysis**

Despite the fact that the ferromagnetic component is small, it is clearly not zero. It is not possible to account for it in the $\Gamma_1$ representation of the $P6_3cm$ magnetic group, as assumed by some authors [10]. Indeed, $\Gamma_1$ is represented only once in this group and the associated symmetry vector correspond to the in-plane antiferromagnetic order. In the magnetic $P6'_3$ group, however the $\Gamma_4$ irreductible representation is represented three times, the ferromagnetic component being the order parameter associated with the **$V_3$** symmetry vector (see fig. 4).

Let us now go back to the Landau analysis and assume that the origin of the ferromagnetic component M comes from the Dzyaloshinskii–Moriya coupling.

$$F = a_2(T-T_N)(A^2+B^2) + a_4(A^2+B^2)^2 - \alpha_2(P^2-P_0^2) + \alpha_4(P^4-P_0^4)$$
$$+ \gamma (S^{ab})^2(P^2-P_0^2) + b_2M^2 + D\mathbf{A}\cdot\mathbf{M}$$

where D is the Dzyaloshinskii–Moriya constant and $b_2M^2$ the ferromagnetic energy. Let us assume in addition that D can be expended in power of the atomic displacements. These displacements can be expressed as a function of only the c axis polarization P. The first symmetry allowed term is thus $P^2-P_0^2$ since $P=P_0$ at and above $T_N$.

$$F = a_2(T-T_N)(S^{ab})^2 + a_4(S^{ab})^4 - \alpha_2(P^2-P_0^2) + \alpha_4(P^4-P_0^4)$$
$$+ \gamma (S^{ab})^2(P^2-P_0^2) + b_2 M^2 + \beta (P^2-P_0^2) S^{ab} M \cos\phi$$

$\partial F/\partial \phi = 0$ leads to $\phi=0$, that is $(S^{ab})^2 = A^2$

$\partial F/\partial \mathbf{M} = 0$ leads to $\mathbf{M} = - \mathbf{A} (\beta/2b_2)(P^2-P_0^2)$

The combination of $\partial F/\partial S^{ab} = 0$ and $\partial F/\partial P = 0$ leads in a first order approximation to the same equations as previously. On thus get

$$M = P_0^2 (\beta \gamma/2\alpha_2 b_2) [4a_4\alpha_4 - \gamma^2]^{-3/2} (2\alpha_4 a_2 t)^{3/2}$$

The solution $\phi=0$ explains why B=0 and A is the only component observed by neutron scattering.

The equation relating **M** to **A** and $(P^2-P_0^2)$ is in perfect agreement with what is experimentally observed: **M** is much smaller than **A** in amplitude (it scales as $A^3$) and along the **c** axis. The presence of the Dzyaloshinskii–Moriya coupling in the free energy explains the existence of a small ferromagnetic component along the c axis.

The most important consequence of the $\beta (P-P_0)^2 \mathbf{A}.\mathbf{M}$ coupling is that one cannot switch the direction of any of the magnetic orders, clockwise vs counter clockwise rotation of the antiferromagnetic order (sign of **A**) or magnetization (**M**), by switching **P**. Indeed, the polarization can appears only as a square in the $P6'_3$ group. **A** and **M** however are switched simultaneously. This consequence of the Landau analysis explains why it is so difficult to switch the different domains in $YMnO_3$ in the polarization measurements as it was recently pointed out [9]. In terms of possible applications, this type of multiferroics is unfortunately not very useful.

**Conclusion**

Careful measurements of the macroscopic properties of a single crystal of $YMnO_3$ induced the existence in this compound of four order parameters at the magnetic transition: the in-plane antiferromagnetic order (two components, one being zero), the polarization along the c axis, and the ferromagnetic order along c axis. The four parameters present different temperature dependence and different amplitudes. The in-plane antiferromagnetic order is the largest and follows a typical temperature dependence of second order magnetic phase transitions. $P-P_0$ is smaller and linear in t, M finally is the smallest and of the highest order in t. This is one of the most complex reported transitions, as far as we know, as it presents two levels of induced ("secondary") order parameters, namely the polarization variation and the magnetization.

[1] Yakel, H. L.,Koehler,W. C., Bertaut, E. F. & Forrat, E. F. Acta Crystallogr. 16, 957 (1963).
[2] Smolenskii, G. A. & Bokov,V. A. J. Appl. Phys. 35, 915 (1964).
[3] I. G. Ismailzade and S. A. Kizhaev, Sov. Phys. Solid State 7, 236, (1965) ; K. Lukaszewicz and J. Karut-Kalinci´nska, Ferroelectrics 7, 81 (1974).
[4] A. S. Gibbs, K. S. Knight and P. Lightfoot, Phys. Rev. B 83, 094111 (2011).
[5] T. Katsufuji, M. Masaki, A. Machida, M. Moritomo, K. Kato, E. Nishibori, M. Takata, M. Sakata, K. Ohoyama, K. Kitazawa and H. Takagi, Phys. Rev. B 66, 134434 (2002)
[6] B. B. Van Aken, T. M. Palstra, A. Filippetti and N. A. Spaldin, Nature Materials 3 164 (2004).
[7] S. C. Abraham, Acta Cryst. B65, 450 (2009).
[8] A. S. Gibbs, K. S. Knight and P. Lightfoot, Phys. Rev. B 83, 094111 (2011).
[9] T. Choi et al, nature materials 9, 253 (2010).
[10] A. Muñoz, et al. , Phys. Rev. B 62, 9498–9510 (2000).
[11] Seongsu Lee, A. Pirogov, Jung Hoon Han, J.-G. Park, A. Hoshikawa, and T. Kamiyama, Phys. Rev. B 71, 180413 (2005).
[12] P. J. Brown, T. Chatterji T, J. of physics : cond . matter 18, 10085 (2006).
[13] E. F. Bertaud, R. Pauthenet et M. Mercier, Phys. Lett. 7, 110 (1963).
[14] S. Pailhès, X. Fabrèges, L. P. Régnault, L. Pinsard-Godart, I. Mirebeau, F. Moussa, M. Hennion, and S. Petit, Phys. Rev. B 79, 134409 (2009).
[15] S. Lee et al., nature letters 451, 805 (2008).
[16] M. Fiebig, Th. Lottermoser, D. Frohlich, A. V. Goltsev, R. V. Pisarev, nature 419, 818 (2002).
[17] D. I. Bilc, R. Orlando, R. Shaltaf, G.-M. Rignanese, Jorge Íñiguez and Ph. Ghosez, Phys. Rev. B 77, 165107 (2008).
[18] S. H. Kim, S. H. Lee, T. H. Kim, T. Zyung, Y. H. Jeong, and M. S. Jang. Growth, Crys. Res. Tech. 35, 19 (2000).
[19] S.W. Cheong, M. Mostovoy, Nature Materials 6, 13 (2007)